\documentclass[openbib,a4paper,12pt]{article}
\usepackage{epsfig}
\begin{document}
\setcounter{section}{0}
\setcounter{equation}{0}
\noindent 
\begin{center}
{\bf \Large The low-energy constants of the $\pi N$ system}
\end{center}

\vspace{1.0cm}
\noindent 
\begin{center}
E.~Matsinos$^\dagger$
\end{center}

\vspace{0.3cm}
\noindent
\begin{center}
Institute for Theoretical Physics, University of Zurich, \\
Winterthurerstrasse 190, CH-8057 Zurich, Switzerland \\
\end{center}

\noindent 
\vspace{0.7cm}
\begin{center}
{\bf Abstract}
\end{center}

\vspace{0.3cm}
\noindent 
Recent analyses of low-energy $\pi N$ experimental data have provided clear evidence for the violation of the isospin symmetry of 
the strong interaction. In the present work, it is shown that the single-charge-exchange reaction ($\pi^- p \rightarrow \pi^0 n$) 
is the culprit for the effect. Given the present experimental uncertainties, no evidence for isospin breaking was found in the two 
elastic-scattering processes for pion laboratory kinetic energies between $20$ and $100$~$MeV$. In agreement with most of the 
recent determinations, the value for the charged-pion coupling constant $f_{\pi^\pm p n}$, extracted herein, is `small'. The energy 
dependence of the $s$- and $p$-wave hadronic phase shifts, obtained from low-energy elastic-scattering data exclusively, as well as
their values in tabular form (including meaningful uncertainties), are provided. Discrepancies with two `standard' phase-shift 
solutions in the $s$-wave part of the interaction are seen; small differences may be observed in three of the $p$-wave channels. 

\vspace{1cm}
\noindent {\it PACS numbers:} 13.75.Gx, 11.30.Hv, 25.80.Dj

\vspace{0.7cm}
\noindent $^\dagger$ Electronic mail: evangelos.matsinos@psi.ch.

\newpage
\section{Introduction}
\noindent 
All pion-nucleon ($\pi N$) measurements (differential cross sections, partial-total cross sections, total nuclear cross sections, 
and analyzing powers) in the three low-energy experimentally-accessible channels~\footnote{The low-energy $\pi N$ channels, which 
are susceptible to experimentation (at present), are the two elastic-scattering processes ($\pi^\pm p$) and the 
single-charge-exchange (SCX) reaction ($\pi^- p \rightarrow \pi^0 n$).} between (pion laboratory kinetic energies of) $20$ and 
$100$~$MeV$ were recently analyzed in the framework of a relativistic isospin-symmetric model~\cite{1}; due to the existence of 
discrepant measurements (outliers) in the input data base, robust statistics was implemented in the problem. Provided the 
correctness of the bulk of the experimental data and of the electromagnetic corrections applied to the scattering problem, it was 
concluded (in Ref.~\cite{1}) that the isospin symmetry of the strong interaction is violated in the $\pi N$ system at low energies.
The reproduction of the input data was subsequently investigated~\cite{2}; it was shown that, with the exclusion of about $14$~\% 
of the elastic-scattering measurements (mostly pertaining to the $\pi^+ p$ reaction) and the additional renormalization of 
seven (out of 40) $\pi^+ p$ data sets, the input data base becomes internally consistent. 

The present article completes the research program set forward in Ref.~\cite{1}; its objectives may be summarized as follows: \\
a) After isospin-symmetry violation in the $\pi N$ system (at low energies) has been established, the interest should lie with the 
question of which reaction is responsible for the effect; the identification of the culprit is expected to lead to clearer 
ideas as to which physical processes are involved in the breaking. \\
b) The $\pi N$ model, used in Ref.~\cite{1}, constitutes the means for the extraction of new hadronic phase shifts (or amplitudes) 
from the experimental data on the exclusive basis of the low-energy information. Among others, this is essential since the 
dynamical structure of the $\pi N$ interaction might be energy-dependent; put in different words, it might be that hadronic 
symmetries, obeyed (by the $\pi N$ system) at high energies (above the $\Delta_{33}$ resonance), are not valid close to the $\pi N$ 
threshold (zero kinetic energy of the incident pion). Such an effect could remove some of the acclaimed `discrepancies' between the
theoretical predictions, obtained via dispersion relations mainly on the basis of high-energy data, and the low-energy
measurements. Furthermore, reliable low-energy information (in the form of hadronic phase shifts) is expected to be of great 
interest in calculations conducted within the framework of the Chiral-Perturbation Theory. \\
c) The model, used herein, has already been established as a firm basis for the extraction of low-energy hadronic constants of the 
$\pi N$ system. In turn, such an extraction is essential because: i) The study of the $\pi N$ system yields information on 
constants which play a fundamental role in other research fields, e.g., in the $N N$ sector, where the $\pi N$ interaction 
comprises the microscopic input; the coupling constants, corresponding to the $\pi N N$ and $\pi N \Delta$ vertices, belong to 
this category. ii) In view of the conclusions of Ref.~\cite{1}, the discussion about some of the low-energy hadronic constants 
might only be meaningful if the corresponding input data base is specified. For instance, instead of talking about {\it one} 
coupling constant $g_{\pi N N}$, one might have to distinguish between $g_{\pi^+ p n}$ (deduced from $\pi^+ p$ data), 
$g_{\pi^- p n}$ (deduced from $\pi^- p$ data), and $g_{\pi^0 p p}$ or $g_{\pi^0 n n}$ (involved in the SCX reaction). Therefore, 
it comes as a natural consequence of the findings of Ref.~\cite{1} to investigate which of the model parameters are affected by 
the effect reported therein. iii) A legitimate question relates to the physical processes involved in the observed 
isospin-breaking effect; their identification might be enabled with the proper comparison of the various parameter values obtained 
during the fitting procedure~\footnote{In Ref.~\cite{1}, the measurements in the three low-energy reaction channels were analyzed 
separately leading to three classes of parameter values. A combined fit to the elastic-scattering data yielded a fourth class.}. \\
In the context of the present work, two kinds of possible differences in the values of the model parameters are relevant. i)
Differences between the results obtained in the single-channel analyses of the measurements in the two elastic-scattering 
reactions.
In the following, these differences will constitute `type I' effects; they are expected to create isospin breaking in the two 
elastic-scattering processes. ii) Differences between the results of the combined fits to the elastic-scattering measurements 
and the ones obtained from the fits to the SCX data. In the following, these differences are to be referred to as `type II' 
effects.

It should be reminded that the model parameters (determined from the various fits to the low-energy $\pi N$ measurements) are: \\
a) $G_\sigma$ and $\kappa_\sigma$ for the scalar-isoscalar $t$-channel exchange, \\
b) $G_\rho$ and $\kappa_\rho$ for the vector-isovector $t$-channel exchange, \\
c) $g_{\pi N N}$ and $x$, standing for the $\pi N N$ coupling constant and the pseudoscalar admixture in the $\pi N N$ vertex,
respectively, and \\
d) $g_{\pi N \Delta}$ and $Z$, the former denoting the $\pi N \Delta$ coupling constant, the latter being associated with the 
spin$-\frac{1}{2}$ admixture in the $\Delta$-isobar field. \\
It was observed in Ref.~\cite{3} that the analysis of low-energy $\pi N$ data cannot lead to the determination of the parameter 
$G_\rho$. For an unbiased analysis, this parameter was fixed at seven (equidistant) values between $30$ and $60$~$GeV^{-2}$; the 
interval chosen corresponds to the extreme cases found in the literature. All details about the $\pi N$ interaction model, used
herein, may be found in Refs.~\cite{1} and~\cite{3}. The method of the analysis has been thoroughly described in Ref.~\cite{1}.
With one exception (to be cited in Section 3), the input data base has been listed in Ref.~\cite{1}.

\section{The two elastic-scattering reactions}
\subsection{On isospin-symmetry violation}
\noindent
Firstly, type I effects were investigated. The following steps were carried out: \\
I. Combined fits to all low-energy elastic-scattering data were performed. From these fits, Solution A (for the seven 
model parameters) was obtained. A strong dependence of the parameters $G_\sigma$, $\kappa_\rho$, and $x$ on $G_\rho$ was 
observed in agreement with the findings of Ref.~\cite{3}. \\
II. Subsequently, {\it separate} fits to the data in the two elastic-scattering reactions were to be performed. However, it is
known (from Ref.~\cite{1}) that a seven-parameter (exclusive) fit to the $\pi^+ p$ measurements cannot be carried out because of 
the problem of the large correlations among the model parameters. Therefore, it was decided to fix some of the model parameters at 
the corresponding values of Solution A. Evidently, a question arises: Which of the model parameters are less likely to be 
influenced by a potential violation of isospin symmetry in the elastic scattering, and, therefore, may be fixed? A clue may be 
obtained by recalling the fact that the $\pi N$ interaction is so weak at low energies (as a consequence of the underlying 
approximate chiral symmetry) that the tree-level version of the model provides a satisfactory description of the $\pi N$ dynamics 
up to about $40$~$MeV$ (see Ref.~\cite{3}). At the tree level, no isospin-breaking effect is expected in the scalar-isoscalar 
$t-$channel graph; hence, $G_\sigma$ and $\kappa_\sigma$ may be fixed. Additionally, the coupling constant $g_{\pi N N}$, obtained 
from $\pi^+ p$ data, corresponds to the $\pi^+ p n$ vertex, whereas the one, resulting from the $\pi^- p$ fits, relates to 
$\pi^- p n$; due to charge symmetry (which is a looser constraint than isospin symmetry), these two coupling constants are expected
to be equal. An additional assumption may be that all the $\Delta-$isobar states involve the same parameter $Z$. Therefore, it 
might be concluded that $\kappa_\rho$, $x$, and $g_{\pi N \Delta}$ could be chosen as free parameters in these fits. If the 
analysis is carried out in this manner, effects in $\kappa_\rho$ and $g_{\pi N \Delta}$ are observed (i.e., these two parameters 
come out different in the two types of fits). \\
III. In order to establish the aforementioned effect (in $\kappa_\rho$ and $g_{\pi N \Delta}$), the influence of the outliers has 
to be carefully examined. Although it is true that, in general, the results from a robust fit are not sensitive to the treatment of
the outliers, there exist three reasons why such an investigation is necessary: a) Most of the discrepant data are contained in the
$\pi^+ p$ reaction (which determines the isospin$-\frac{3}{2}$ amplitudes almost exclusively). b) The distribution of the 
normalized residual $z$ of the $\pi^+ p$ data is asymmetrical (see Fig.~3(a) in Ref.~\cite{1}). c) The outliers comprise a 
significant amount of the whole $\pi^+ p$ data base; this is an important observation, especially so when combined with the fact 
that all these measurements lie systematically above the bulk of the data (i.e., they populate the lower tail of the $z$ 
distribution). Due to these reasons, an investigation of the stability of the solutions, obtained at step II, with respect to the 
treatment of the discrepant data is pertinent. Repeating the analysis after the data of Bertin {\it et al.}~\cite{4} and of Carter 
{\it et al.}~\cite{5} were removed from the data base~\footnote{The conclusion of Ref.~\cite{2} is that these data are way off 
any other low-energy $\pi^+ p$ measurement.} showed that the aforementioned effect in $\kappa_\rho$ and $g_{\pi N \Delta}$ was 
spurious; it was simply an artifact of the discrepant data. No statistically significant effects in these two parameters were 
observed after the data of Refs.~\cite{4} and~\cite{5} were excluded. In practice, this implies that both elastic-scattering 
processes may be accounted for by a common set of parameter values. The solution, obtained after the two aforementioned data sets
were excluded from the input data base, will be referred to as \underline{Solution B} and will be exclusively used hereafter. \\
The first conclusion of this work is that, {\it in the energy region, dealt with herein, and given the present experimental 
uncertainties}, there is no evidence for isospin breaking in the two elastic-scattering reactions. This observation is consonant
with Weinberg's statement that isospin breaking in the $\pi N$ system involves (at least) one neutral pion~\cite{6}.

At this point, one comment is prompt. The $\rho^0 - \omega$ mixing mechanism~\cite{7}-\cite{8} was proposed in the past as a 
potential means for isospin breaking in $\pi N$ elastic scattering. Despite the fact that such effects have not been seen in
the energy region of the present work, one should remark that (possible) deviations from isospin symmetry may be observed at 
places where the main parts of the interaction (i.e., the real parts of the $s$- and $p$-wave amplitudes) cancel each other; at 
these places, small effects are magnified. One of these regions was recently explored experimentally in Ref.~\cite{9}. An analysis 
of more recent data, taken at the same kinematic region and accompanied by a significant improvement in the experimental 
systematic uncertainties, is expected to shed more light on this matter.

\subsection{Low-energy information extracted from elastic-scattering data}
\noindent
Let us start with a discussion on each one of the model parameters. The results are concisely contained in Figs.~1 and~2
which show the $G_\rho$ dependence of the model parameters fitted to the data in the optimization phase. The 
three values displayed (per $G_\rho$ point) correspond to the cases where the (small) $d$ and $f$ waves: a) were fixed at the 
values of the most recent Karlsruhe solution (KA85)~\cite{10}, b) were fixed at the values of the VPI solution SM95~\cite{11}, and 
c) were calculated with the $\pi N$ interaction model~\footnote{It should be kept in mind that the $s$- and $u$-channel 
contributions of the $d$ and $f$ (baryon) resonances (to the $K$-matrix elements of the model) cannot be calculated due to 
the problem relating to the quantization of high-spin fields.}. In all cases, the agreement among these three solutions is very 
satisfactory. \\
a) {\it $G_\sigma$} is a decreasing function of $G_\rho$; it varies between (approximately) $41$ and $27$~$GeV^{-2}$. Its large 
values signify the importance of the scalar-isoscalar interaction in the description of the low-energy $\pi N$ dynamics. \\
b) {\it $\kappa_\sigma$} is found to be almost independent of $G_\rho$; its average value is about two to three standard deviations
away from zero. The smallness of this value is the reason why, prior to Ref.~\cite{1} (i.e., prior to the introduction of the 
derivative coupling of the scalar-isoscalar to the pion field in the model), the $\pi N$ data could also be accounted for by the 
model very successfully. \\
c) {\it $\kappa_\rho$} is a decreasing function of $G_\rho$. In general, the values, obtained herein, are significantly smaller 
than the ones extracted with dispersion-relation techniques (which lie in the vicinity of $6$, e.g., see Ref.~\cite{12}); the 
vector-meson-dominance prediction ($3.7$) is in agreement with the values of the present work, in the lower part of the interval 
corresponding to the $G_\rho$ variation. Brown and Machleidt~\cite{13} have argued that instead of comparing the $G_\rho$ and 
the $\kappa_\rho$ values in different analyses, one should rather quantify the overall strength of the $\rho$ coupling as
\begin{center}
$(g_{\rho N N}^{(V)} )^2$ \ $\frac{1+\kappa_\rho^2}{4 \pi}$ \ \ \ ,
\end{center}
where $g_{\rho N N}^{(V)}$ is the vector coupling constant corresponding to the $\rho N N$ vertex. In this context, the value of
the present work for the overall strength of the $\rho$ coupling (which is around $9$) supports the weak-$\rho$ scenario; this 
result is in sharp contradiction to the conclusions of analyses achieved with dispersion-relation techniques. \\
d) {\it $g_{\pi N N}$} is probably the hadronic constant with the longest history, a fact signifying its importance also outside 
the field of Pion Physics. A compilation of pre-meson-factory values may be found in Refs.~\cite{14}. The value of $g_{\pi N N}$, 
extracted in Ref.~\cite{15}, was found to reproduce the $\pi N$ experimental data in the Karlsruhe-Helsinki (KH80) 
analysis~\cite{16} which was simultaneous to the first results from meson-factory experiments. The most recent literature values 
for $g_{\pi N N}$ are listed in Table~1 along with the values extracted in this work. One should remark that these values 
originate from analyses of $\pi N$ (Refs.~\cite{16}, \cite{17}, \cite{18}, \cite{19}, and the present work), as well as $N N$ and 
$\overline{N} N$ data (the outcome of a long history of research from the Nijmegen group (e.g., see Refs.~\cite{20} and the 
articles cited therein), and the values of Refs.~\cite{21} and~\cite{22}). The result of Ref.~\cite{23} has not been included in 
the table because it was recently revised~\cite{24} and the new value is still considered to be preliminary~\cite{25}; this value 
is in agreement with Ref.~\cite{16}. To enable the easy comparison of the various items found in the literature, the values of the
strong coupling constant are given in Table~1 in the form:
\begin{equation}
f_{\pi N N}^2 \ = \ (\frac{m_\pi}{2m})^2 \ \frac{g_{\pi N N}^2}{4 \pi} \ \ \ ,
\end{equation}
where $m_\pi$ and $m$ denote the charged-pion and the proton mass, respectively. With the exception of the values of 
Refs.~\cite{16} and~\cite{22}, general agreement is observed among the entries of Table~1. \\
e) {\it x} is found to be $G_\rho$-dependent. In all cases, the pseudoscalar admixture in the $\pi N N$ vertex is small. \\
f) {\it $g_{\pi N \Delta}$}. The values of the present work are compatible with determinations based on the $\Delta$-isobar width 
(e.g., see Ref.~\cite{26}) and are totally inconsistent with the dispersion-relation result~\cite{27}. \\
g) {\it Z} has an average value of about $-0.39$. This value is compatible with the earlier determination of Ref.~\cite{3} and 
signifies the importance of the proper relativistic treatment of the $\Delta$-isobar.

Figures~3 and~4 show the energy dependence of the $s$- and $p$-wave hadronic phase shifts corresponding to the combined fits to the 
elastic-scattering data; the values represent averages over the three options for the $d$ and $f$ waves (the sensitivity of the 
results on the choice of the $d$ and $f$ waves was checked and found to be small compared to the statistical uncertainties). To 
enable the straightforward application of the results of the present analysis, the $s$- and $p$-wave hadronic phase shifts in the 
low-energy domain are listed in Tables~2. The conclusions, drawn on the basis of Figs.~3 and~4 and Tables~2, may be summarized as 
follows: \\ 
a) The KA85 solution is incompatible with our results in the $s$-wave part of the interaction. Although, the SM95 solution does a 
better job, there is definitely some discrepancy (with our solution) in the $S_{31}$ phase shift (the disagreement being milder 
than in the case of the KA85). It has to be noted that the $S_{31}$ values, extracted in the present work, are in excellent 
agreement with the results of Ref.~\cite{28} which were exclusively based on the recent $\pi^+ p$ differential-cross-section 
measurements. \\
b) Although the overall status in the $p$-wave part of the interaction seems to be satisfactory, a difference of about one degree
between our results and the KA85 values was detected in the $P_{33}$ channel (close to the highest energy allowed in the present 
work); our solution is (once more) in excellent agreement with the values of Ref.~\cite{28}. All solutions agree well in the 
$P_{31}$ channel. A small, yet systematic, deviation from the SM95 $P_{13}$ phase-shift values was observed. Finally, there is a
good agreement between our values in the $P_{11}$ channel and the SM95 solution, whereas, in this channel, the KA85 values are 
smaller by half a degree at (pion center-of-mass kinetic energy of) $60$~$MeV$.

Finally, one should state that our results for the $p$-wave part of the interaction are in disagreement with the recent findings 
of Ref.~\cite{19}. The present article cannot support the statement of Ref.~\cite{19} that the hadronic phase shifts (and the 
corresponding scattering volumes) in the $P_{31}$ and $P_{13}$ channels are equal (see Figs.~3 and~4 and Tables~2). Additionally, 
we do not find a smaller value (e.g., than the value of Ref.~\cite{10}) for the scattering volume in the $P_{33}$ channel.

\section{The SCX reaction}
\noindent
Based on the conclusions, drawn from the analysis of the measurements in the two elastic-scattering processes, one has to attribute
the isospin-breaking effect, reported in Ref.~\cite{1}, to the SCX reaction. Isospin breaking may enter the interaction in two 
ways; either as contributions from (isospin-breaking) diagrams which are not present in the model (external breaking) or as a 
difference in some of the coupling constants and/or vertex factors in the Feynman diagrams of the model (internal breaking). In 
Ref.~\cite{1}, it was found that the model can account for the SCX reaction successfully. In practice, this observation implies
that, if the breaking is external, the model contributions can successfully mimic the missing pieces. Since we do not
possess a way to distinguish between internal and external effects, the parameter values, obtained from the fits to the SCX data, 
will not be taken too seriously though they do come out reasonable from the fits; only the amplitude (which mainly depends on the 
experimental data and the electromagnetic corrections) will be assigned significance. Five-parameter fits to the SCX data were 
carried out~\footnote{Notice that, as in Ref.~\cite{1}, the scalar-isoscalar parameters were fixed at zero for the fits to the SCX 
data on the basis of the validity of the tree-level approximation at low energies; in such a case, the scalar-isoscalar interaction
cannot produce a neutral pion in the final state (starting from a charged projectile).} and the SCX amplitude was successfully 
constructed. The experimental data of Ref.~\cite{29}, which were recently finalized, were also included in the data base; these
measurements had not been available at the time the analysis of Ref.~\cite{1} was completed.

Table~3 shows the isospin-breaking effect in the $s$-wave part of the interaction (the effects in the $p$-wave component are not
statistically significant) for three cases, corresponding to the three treatments of the $d$ and $f$ waves, in the form of the 
symmetrized ratio:
\begin{equation}
ISB \ = \ 2 \ \frac{Re f_{SCX} \ - \ Re f_{SCX}^{\pi^\pm p}}{Re f_{SCX} \ + \ Re f_{SCX}^{\pi^\pm p}} \ \ \ ;
\end{equation}
$f_{SCX}$ denotes the SCX amplitude extracted directly from SCX data and $f_{SCX}^{\pi^\pm p}$ stands for the SCX amplitude as
predicted from the elastic-scattering data. (Evidently, in case that isospin symmetry holds, $ISB$ should equal $0$.)
An energy dependence of $ISB$ was not observed. The effect is large (about $7.5$~\%) and statistically significant.

It was found that the model parameters $x$ and $Z$ come out different in elastic scattering and SCX; as seen in Fig.~5, the 
values, extracted from the fits to the SCX measurements, are systematically larger for both parameters. This implies that the 
$s$- and $u$-channel contributions are different in these two cases; the difference in $x$ affects the nucleon graphs, the one in 
$Z$ influences the graphs with a $\Delta$-isobar in the intermediate state. Of course, it is evident
that the $\rho N N$ vertex might also be different in elastic scattering and in SCX~\cite{30}; since the fits are performed at 
fixed $G_\rho$ values, such a difference is expected to be transferred to $x$ (since $x$ is correlated with $G_\rho$). 
Unfortunately, one cannot distinguish between these internal effects. Finally, the question arises as to which role external 
effects (e.g., the $\eta - \pi^0$ mixing mechanism~\cite{8}-\cite{31}) play in this issue. This mechanism is expected to affect 
all channels; on top of the straightforward $s$- and $u$-channel modifications due to this mechanism, a Feynman diagram with an 
$a_0 (980)$ $t$-channel exchange might make large contributions since the dominant decay of $a_0 (980)$ is in the $\eta \pi$ mode. 
An estimation of these effects is needed before one associates certain physical processes with the isospin breaking established in 
the low-energy $\pi N$ scattering.

\section{Conclusions}
\noindent
One of the aims of the present work was the identification of the reaction creating the isospin breaking~\cite{1} 
in the $\pi N$ system at low energies (pion laboratory kinetic energy between $20$ and $100$~$MeV$); the SCX reaction is the 
culprit for the effect. The source of the breaking may involve a difference in some coupling constants and/or vertex factors in 
the Feynman diagrams of the model or be exclusively due to missing (isospin-breaking) pieces. An estimation of the contributions, 
which are not included in the model, is necessary in order to associate certain physical processes with the effect.

In the energy region, dealt with herein, and given the present experimental uncertainties, there is no evidence for 
isospin breaking in the two elastic-scattering processes. The possibility of additional investigation of this issue,
at kinematical regions where the main parts of the interaction cancel each other, is left open; such an investigation may be
enabled after the finalization of the data of an improved version of the experiment of Ref.~\cite{9}.

The model parameters, obtained from the combined fits to the elastic-scattering data, have been given (Figs.~1 and~2). The values 
of the charged-pion coupling constant, extracted herein, were found to be `small' (Table~1). The energy dependence of the $s$- and 
$p$-wave hadronic phase shifts, obtained from the combined fits to the elastic-scattering data, has been provided (Figs.~3 and~4) 
and the corresponding values have been listed (Tables~2) ready for use since meaningful uncertainties are also quoted. It was 
found that the KA85 solution~\cite{10} is incompatible with our results in the $s$-wave part of the interaction. Although, the SM95 
solution~\cite{11} does a better job, there is definitely some discrepancy (with our solution) in the $S_{31}$ phase shift. Small
differences (among the solutions) may be observed in three $p$-wave channels. The present analysis cannot support 
the statements of Ref.~\cite{19} concerning the $p$-wave component of the interaction.

The stability of the aforementioned results on the treatment of the $d$ and $f$ waves has been investigated; three 
sources for their values (the model, Ref.~\cite{10}, and Ref.~\cite{11}) have been assumed, leaving the results of this work 
intact. Finally, it should be mentioned that all the above results rely on the correctness of the bulk of the existing low-energy 
$\pi N$ experimental data. It is also assumed that the electromagnetic corrections of the NORDITA group~\cite{32}, which have been
exclusively used in this research program, are not largely erroneous.
\\~\\~\\~\\
{\bf Acknowledgements.} 
I would like to thank A.B. Gridnev for several edifying discussions and for drawing my attention to the significance of the 
contributions of the $d$ and $f$ waves at some kinematical regions in the $\pi N$ system. I acknowledge helpful discussions with 
A. Badertscher, P.F.A. Goudsmit, Ch. Hilbes, M. Janousch, H.J. Leisi, and P. Weber.

\newpage
\begin{center}
\begin{tabular}{|c|c|c|c|}
\hline
Reference & Data base & $f_{\pi N N}^2$ & Vertex type \\
\hline
This work (model $d$ and $f$ waves)            & $\pi^\pm p$      & $(76.6 \pm 1.1) 10^{-3}$         & $\pi^\pm p n$ \\
This work ($d$ and $f$ waves of Ref.~\cite{10}) & $\pi^\pm p$      & $(75.5 \pm 1.2) 10^{-3}$         & $\pi^\pm p n$ \\
This work ($d$ and $f$ waves of Ref.~\cite{11}) & $\pi^\pm p$      & $(74.5 \pm 1.2) 10^{-3}$         & $\pi^\pm p n$ \\
\cite{16} & $\pi N$          & $(79 \pm 1) 10^{-3}$             & $\pi N N$ \\
\cite{17} & $\pi N$          & $(77.1 \pm 1.4) 10^{-3}$         & $\pi N N$ \\
\cite{18} & $\pi N$          & $(76 \pm 1) 10^{-3}$             & $\pi N N$ \\
\cite{19} & $\pi^\pm p$      & $(75.6 \pm 0.7) 10^{-3}$         & $\pi^\pm p n$ \\
\cite{20} & $np$             & $(74.8 \pm 0.3) 10^{-3}$         & $\pi^\pm p n$ \\
\cite{20} & $pp$             & $(74.5 \pm 0.6) 10^{-3}$         & $\pi^0 p p$ \\
\cite{20} & $\overline{p}p$  & $(73.2 \pm 1.1) 10^{-3}$         & $\pi^\pm p n$ \\
\cite{21} & $np$             & $(75.7 \pm 0.8 \pm 1.3) 10^{-3}$ & $\pi^\pm p n$ \\
\cite{21} & $pp$             & $(77.1 \pm 0.9 \pm 0.4) 10^{-3}$ & $\pi^0 p p$ \\
\cite{22} & $\overline{p}p$  & $(71 \pm 2) 10^{-3}$             & $\pi^\pm p n$ \\
\hline
\end{tabular}
\end{center}
\vspace{1cm}
{\large \bf Table 1:}
The values of the present work for the strong coupling constant as obtained from the combined fits to the elastic-scattering data
along with literature values. 

\newpage
\begin{center}
\begin{tabular}{|c|c|c|c|}
\hline
$T_\pi$ & $S_{31}$ & $P_{33}$ & $P_{31}$ \\
\hline
 $ 20.0$ & $-2.34 \pm 0.03$ & $ 1.30 \pm 0.01$ & $-0.23 \pm 0.01$ \\
 $ 25.0$ & $-2.73 \pm 0.03$ & $ 1.84 \pm 0.01$ & $-0.32 \pm 0.01$ \\
 $ 30.0$ & $-3.12 \pm 0.03$ & $ 2.46 \pm 0.02$ & $-0.41 \pm 0.01$ \\
 $ 35.0$ & $-3.50 \pm 0.03$ & $ 3.16 \pm 0.02$ & $-0.51 \pm 0.02$ \\
 $ 40.0$ & $-3.88 \pm 0.03$ & $ 3.93 \pm 0.02$ & $-0.62 \pm 0.02$ \\
 $ 45.0$ & $-4.27 \pm 0.03$ & $ 4.78 \pm 0.03$ & $-0.73 \pm 0.03$ \\
 $ 50.0$ & $-4.66 \pm 0.04$ & $ 5.72 \pm 0.03$ & $-0.84 \pm 0.03$ \\
 $ 55.0$ & $-5.05 \pm 0.04$ & $ 6.74 \pm 0.03$ & $-0.96 \pm 0.04$ \\
 $ 60.0$ & $-5.45 \pm 0.04$ & $ 7.86 \pm 0.03$ & $-1.08 \pm 0.04$ \\
 $ 65.0$ & $-5.85 \pm 0.05$ & $ 9.09 \pm 0.03$ & $-1.20 \pm 0.05$ \\
 $ 70.0$ & $-6.25 \pm 0.06$ & $10.42 \pm 0.04$ & $-1.32 \pm 0.06$ \\
 $ 75.0$ & $-6.66 \pm 0.06$ & $11.87 \pm 0.04$ & $-1.45 \pm 0.06$ \\
 $ 80.0$ & $-7.07 \pm 0.07$ & $13.46 \pm 0.04$ & $-1.58 \pm 0.07$ \\
 $ 85.0$ & $-7.49 \pm 0.08$ & $15.18 \pm 0.04$ & $-1.71 \pm 0.08$ \\
 $ 90.0$ & $-7.91 \pm 0.09$ & $17.05 \pm 0.05$ & $-1.84 \pm 0.09$ \\
 $ 95.0$ & $-8.34 \pm 0.11$ & $19.09 \pm 0.06$ & $-1.98 \pm 0.09$ \\
 $100.0$ & $-8.76 \pm 0.12$ & $21.31 \pm 0.08$ & $-2.11 \pm 0.10$ \\
\hline
\end{tabular}
\end{center}
\vspace{1cm}
{\large \bf Table 2(a):}
The isospin$-\frac{3}{2}$ $s$- and $p$-wave hadronic phase shifts (in degrees) in the low-energy domain extracted from 
elastic-scattering data. Averages over the three options for the $d$ and $f$ waves (see text) are assumed. $T_\pi$ (in $MeV$) 
denotes the pion laboratory kinetic energy.

\newpage
\begin{center}
\begin{tabular}{|c|c|c|c|}
\hline
$T_\pi$ & $S_{11}$ & $P_{13}$ & $P_{11}$ \\
\hline
 $ 20.0$ & $4.22 \pm 0.02$ & $-0.17 \pm 0.01$ & $-0.36 \pm 0.01$ \\
 $ 25.0$ & $4.70 \pm 0.03$ & $-0.23 \pm 0.01$ & $-0.47 \pm 0.02$ \\
 $ 30.0$ & $5.13 \pm 0.03$ & $-0.30 \pm 0.01$ & $-0.58 \pm 0.02$ \\
 $ 35.0$ & $5.53 \pm 0.03$ & $-0.36 \pm 0.02$ & $-0.68 \pm 0.03$ \\
 $ 40.0$ & $5.88 \pm 0.04$ & $-0.43 \pm 0.02$ & $-0.77 \pm 0.03$ \\
 $ 45.0$ & $6.21 \pm 0.04$ & $-0.50 \pm 0.02$ & $-0.86 \pm 0.04$ \\
 $ 50.0$ & $6.51 \pm 0.05$ & $-0.57 \pm 0.03$ & $-0.93 \pm 0.04$ \\
 $ 55.0$ & $6.78 \pm 0.06$ & $-0.64 \pm 0.03$ & $-0.99 \pm 0.05$ \\
 $ 60.0$ & $7.04 \pm 0.07$ & $-0.71 \pm 0.04$ & $-1.04 \pm 0.05$ \\
 $ 65.0$ & $7.28 \pm 0.08$ & $-0.78 \pm 0.04$ & $-1.07 \pm 0.06$ \\
 $ 70.0$ & $7.50 \pm 0.10$ & $-0.85 \pm 0.05$ & $-1.09 \pm 0.06$ \\
 $ 75.0$ & $7.70 \pm 0.11$ & $-0.92 \pm 0.05$ & $-1.09 \pm 0.07$ \\
 $ 80.0$ & $7.88 \pm 0.13$ & $-0.99 \pm 0.06$ & $-1.07 \pm 0.08$ \\
 $ 85.0$ & $8.05 \pm 0.14$ & $-1.06 \pm 0.06$ & $-1.04 \pm 0.08$ \\
 $ 90.0$ & $8.21 \pm 0.16$ & $-1.12 \pm 0.07$ & $-0.99 \pm 0.09$ \\
 $ 95.0$ & $8.35 \pm 0.18$ & $-1.19 \pm 0.08$ & $-0.92 \pm 0.10$ \\
 $100.0$ & $8.48 \pm 0.20$ & $-1.25 \pm 0.08$ & $-0.83 \pm 0.11$ \\
\hline
\end{tabular}
\end{center}
\vspace{1cm}
{\large \bf Table 2(b):}
The isospin$-\frac{1}{2}$ $s$- and $p$-wave hadronic phase shifts (in degrees) in the low-energy domain extracted from 
elastic-scattering data. Averages over the three options for the $d$ and $f$ waves (see text) are assumed. $T_\pi$ (in $MeV$) 
denotes the pion laboratory kinetic energy.

\newpage
\begin{center}
\begin{tabular}{|c|c|c|c|c|}
\hline
$d$- and $f$-wave treatment & $T_\pi = 30$~$MeV$ &  $T_\pi = 50$~$MeV$ &  $T_\pi = 70$~$MeV$ & Average \\
\hline
From the model      & $8.0 \pm 1.3$~\% & $8.1 \pm 1.2$~\% & $8.2 \pm 1.3$~\% & $8.1 \pm 1.3$~\% \\
From Ref.~\cite{10} & $7.6 \pm 1.3$~\% & $7.6 \pm 1.2$~\% & $7.4 \pm 1.3$~\% & $7.5 \pm 1.3$~\% \\
From Ref.~\cite{11} & $7.4 \pm 1.3$~\% & $7.3 \pm 1.2$~\% & $7.1 \pm 1.3$~\% & $7.2 \pm 1.3$~\% \\
\hline
\end{tabular}
\end{center}
\vspace{1cm}
{\large \bf Table 3:}
The isospin-breaking effect in the $s$-wave part of the interaction for three cases, corresponding to three treatments of the 
$d$ and $f$ waves, in the form of the symmetrized ratio given in Eq.~(2). $T_\pi$ denotes the pion laboratory kinetic energy. The 
entry in the last column represents an average over the three energies of the table which are representative of the input data.

\newpage
\begin{figure}[b]%
\begin{center}%
\epsfig{file=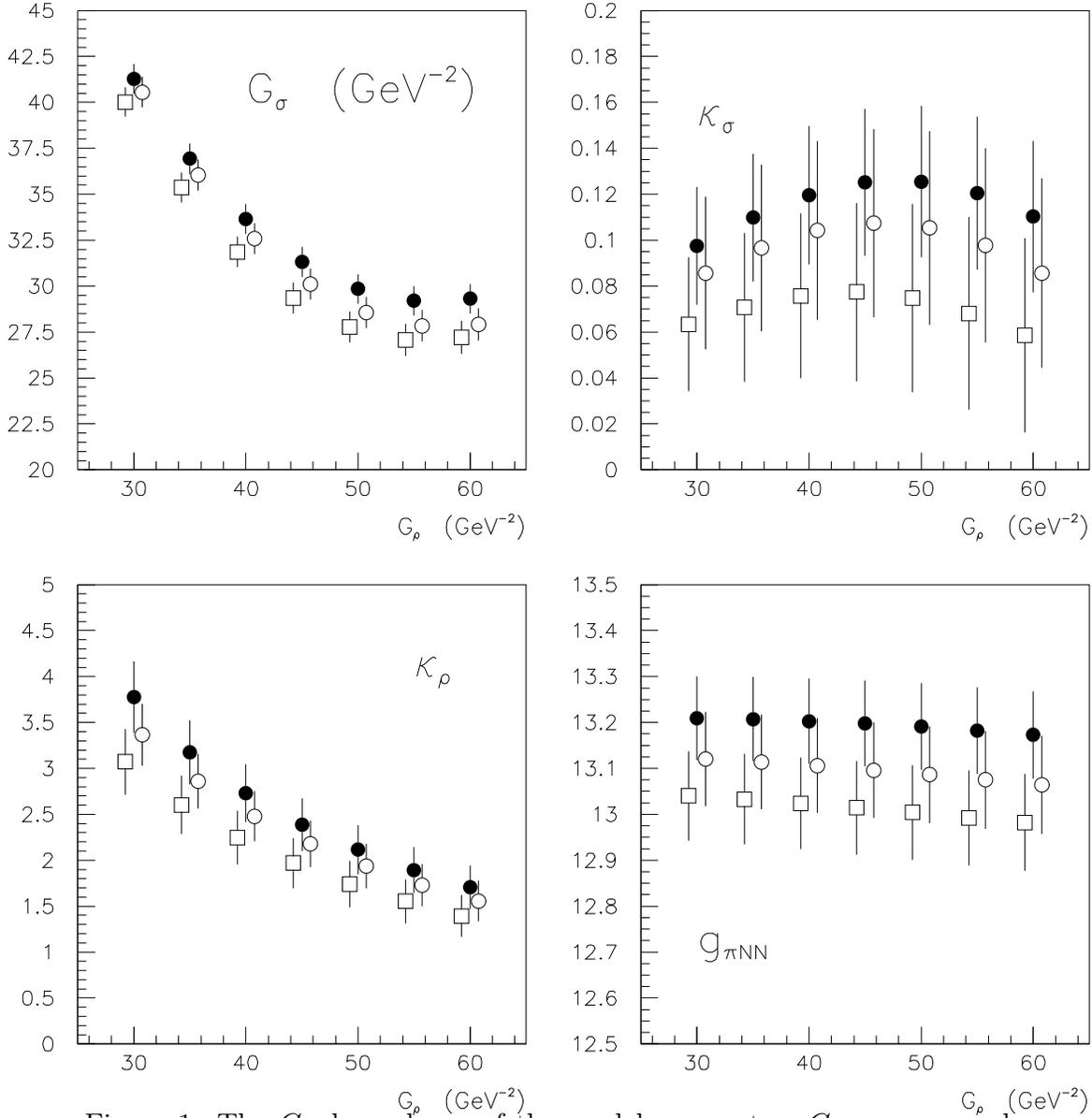,width=15cm,bbllx=2cm,%
        bblly=6cm,bburx=20cm,bbury=28cm,clip=}%
\caption{The $G_\rho$-dependence of the model parameters $G_\sigma$, $\kappa_\sigma$, $\kappa_\rho$, and $g_{\pi N N}$
for the combined fits to the elastic-scattering data and three treatments of the $d$ and $f$ waves; filled circles: the 
$d$ and $f$ waves are calculated with the model, open circles: the $d$ and $f$ waves are taken from Ref.~\cite{10}, squares: the 
$d$ and $f$ waves are taken from Ref.~\cite{11}.}%
\end{center}%
\end{figure}
 
\newpage
\begin{figure}[b]%
\begin{center}%
\epsfig{file=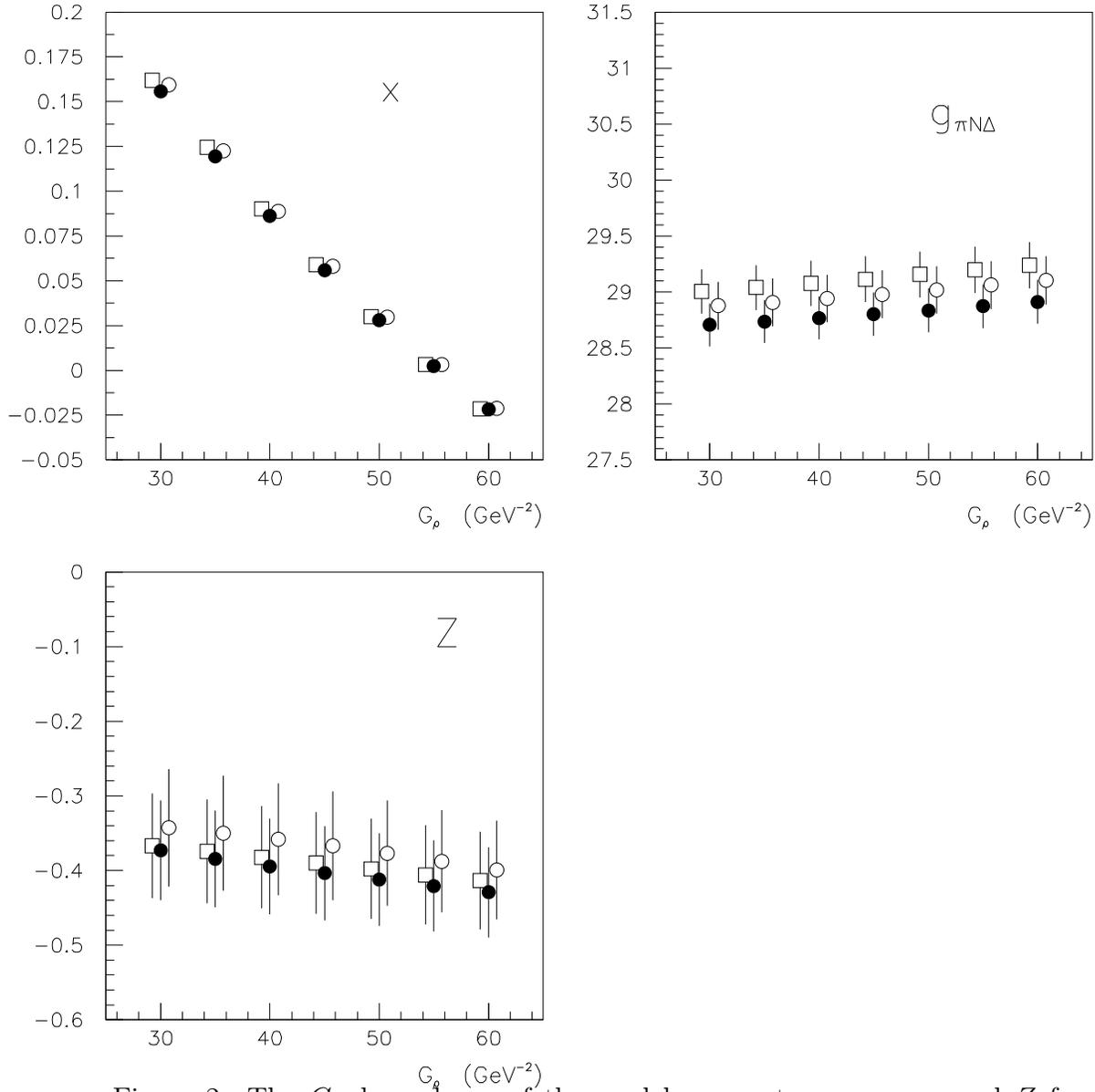,width=15cm,bbllx=2cm,%
        bblly=6cm,bburx=20cm,bbury=28cm,clip=}%
\caption{The $G_\rho$-dependence of the model parameters $x$, $g_{\pi N \Delta}$, and $Z$ for the combined fits to the 
elastic-scattering data and three treatments of the $d$ and $f$ waves; filled circles: the $d$ and $f$ waves are calculated with 
the model, open circles: the $d$ and $f$ waves are taken from Ref.~\cite{10}, squares: the $d$ and $f$ waves are taken from 
Ref.~\cite{11}.}%
\end{center}%
\end{figure}

\newpage
\begin{figure}[b]%
\begin{center}%
\epsfig{file=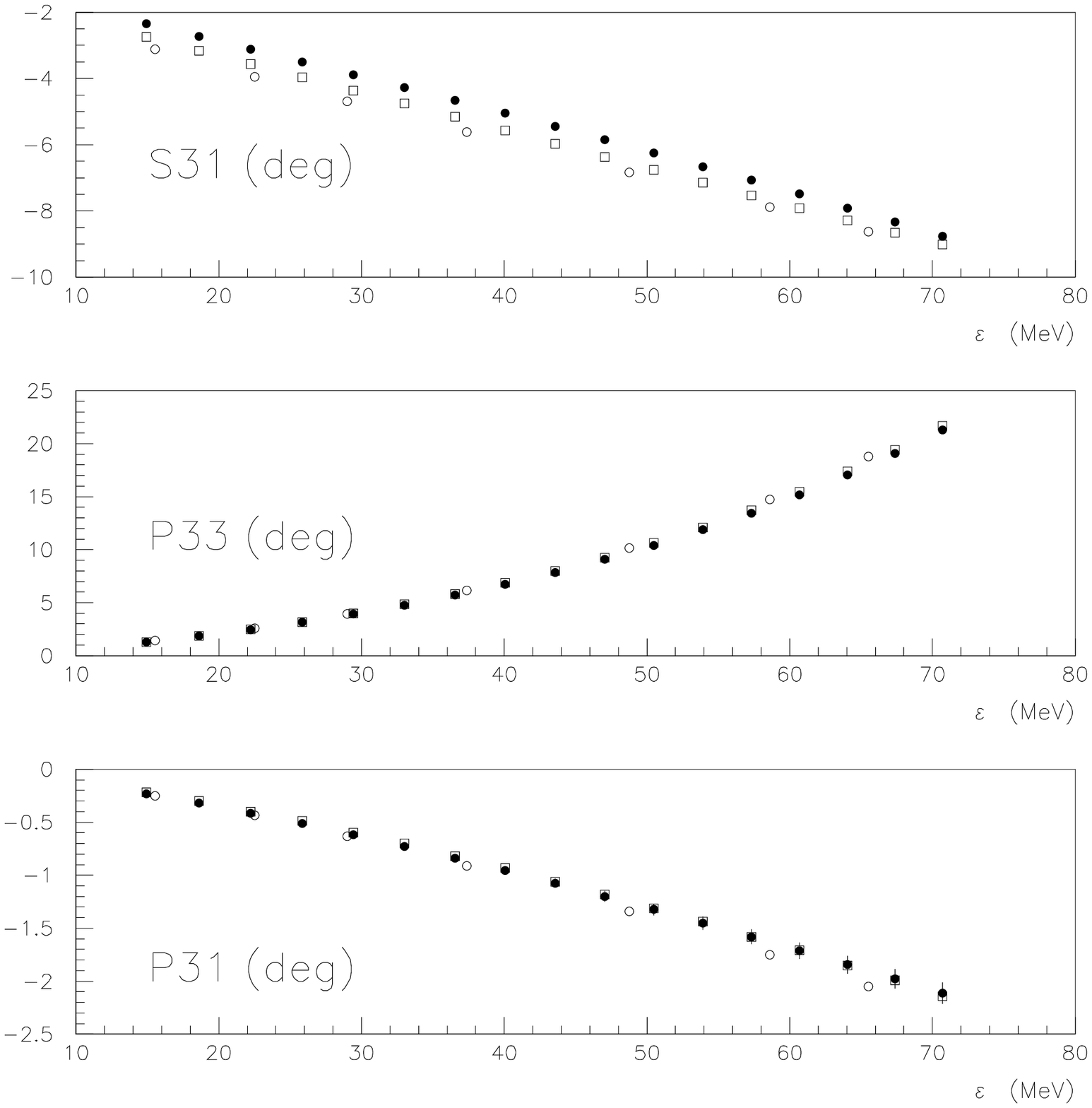,width=15cm,bbllx=2cm,%
        bblly=6cm,bburx=20cm,bbury=28cm,clip=}%
\caption{The isospin$-\frac{3}{2}$ $s$- and $p$-wave hadronic phase shifts corresponding to the combined fits to the 
elastic-scattering data (filled circles); averages over the three options for the $d$ and $f$ waves (see text) are assumed. 
$\epsilon$ stands for the pion center-of-mass kinetic energy. The open circles denote the KA85 solution~\cite{10}, whereas the 
squares represent the SM95 values~\cite{11}.}%
\end{center}%
\end{figure}

\newpage
\begin{figure}[b]%
\begin{center}%
\epsfig{file=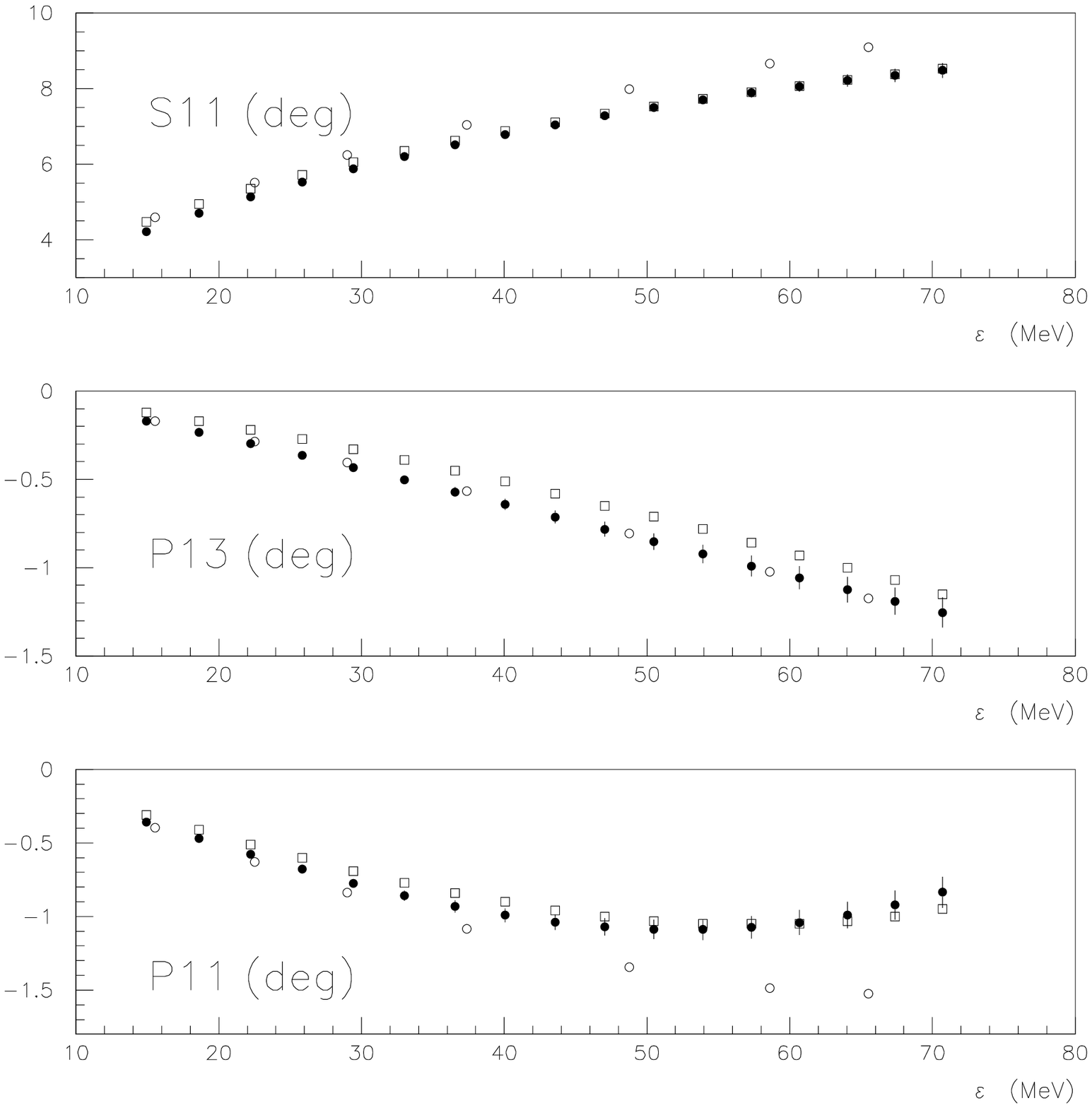,width=15cm,bbllx=2cm,%
        bblly=6cm,bburx=20cm,bbury=28cm,clip=}%
\caption{The isospin$-\frac{1}{2}$ $s$- and $p$-wave hadronic phase shifts corresponding to the combined fits to the 
elastic-scattering data (filled circles); averages over the three options for the $d$ and $f$ waves (see text) are assumed. 
$\epsilon$ stands for the pion center-of-mass kinetic energy. The open circles denote the KA85 solution~\cite{10}, whereas the 
squares represent the SM95 values~\cite{11}.}%
\end{center}%
\end{figure}

\newpage
\begin{figure}[b]%
\begin{center}%
\epsfig{file=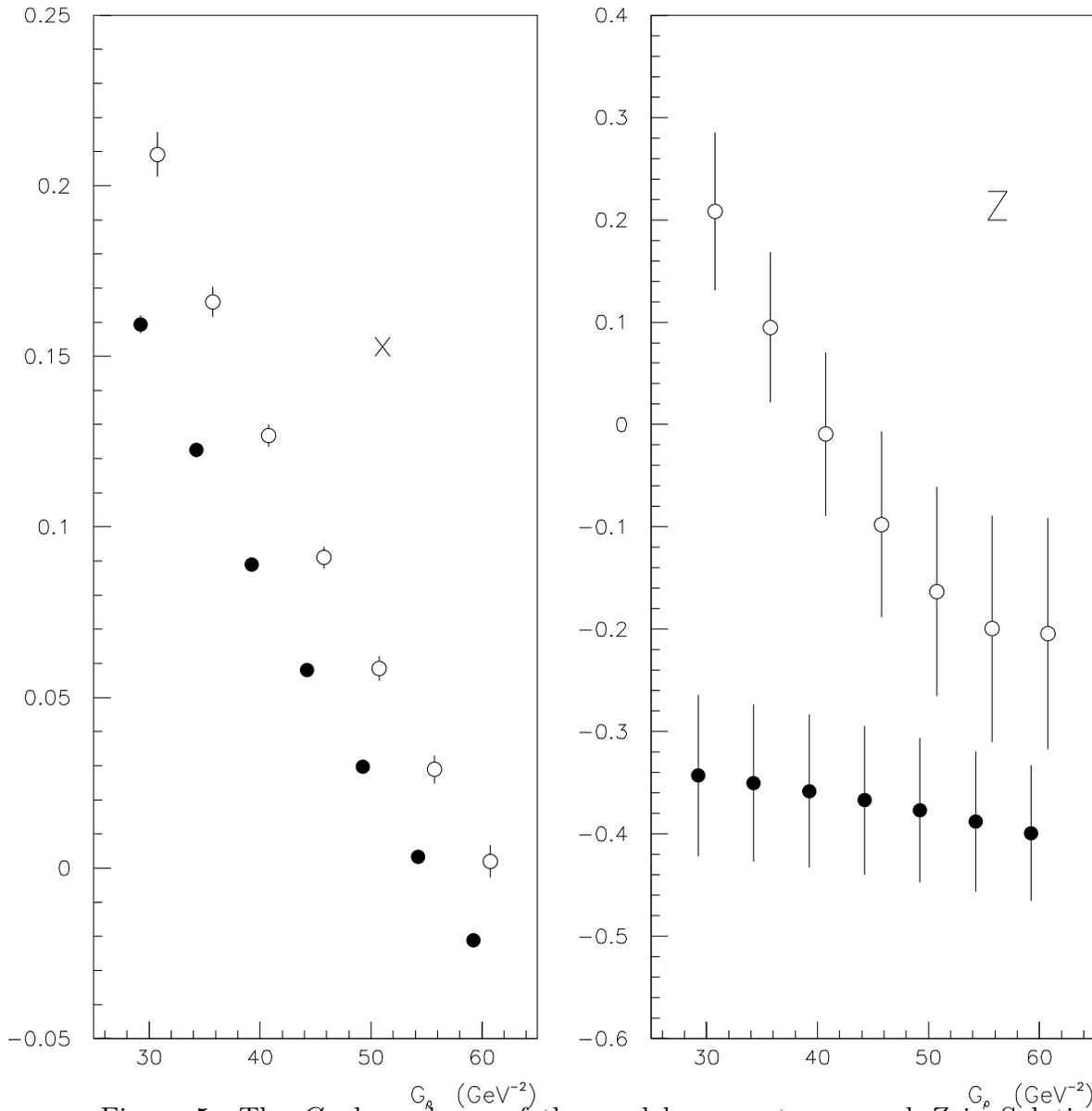,width=15cm,bbllx=2cm,%
        bblly=6cm,bburx=20cm,bbury=28cm,clip=}%
\caption{The $G_\rho$-dependence of the model parameters $x$ and $Z$ in Solution B (see text) and from the fits to the SCX 
measurements (filled and open circles, respectively). The $d$ and $f$ waves have been taken from Ref.~\cite{10}; the other two 
$d$- and $f$-wave treatments (see text) lead to almost identical pictures.}%
\end{center}%
\end{figure}
\end{document}